\documentclass[letterpaper,twocolumn,english,superscriptaddress]{revtex4}
\usepackage[T1]{fontenc}
\usepackage[latin9]{inputenc}
\usepackage{amsmath}
\usepackage{amssymb}
\usepackage{graphicx}
\usepackage{dcolumn}
\usepackage{verbatim}

\makeatletter

\pdfpageheight\paperheight
\pdfpagewidth\paperwidth

\usepackage{babel}

\makeatother

\usepackage{babel}

\usepackage{tikz,xcolor,hyperref}
\definecolor{lime}{HTML}{A6CE39}
\DeclareRobustCommand{\orcidicon}{
	\begin{tikzpicture}
	\draw[lime, fill=lime] (0,0) 
	circle [radius=0.16] 
	node[white] {{\fontfamily{qag}\selectfont \tiny ID}};
	\draw[white, fill=white] (-0.0625,0.095) 
	circle [radius=0.007];
	\end{tikzpicture}
	\hspace{-2mm}
}

\foreach \x in {A, ..., Z}{%
	\expandafter\xdef\csname orcid\x\endcsname{\noexpand\href{https://orcid.org/\csname orcidauthor\x\endcsname}{\noexpand\orcidicon}}
}

\begin{document}

\title{Global spin alignment of vector mesons and strong force fields in heavy-ion collisions}

\author{Jinhui Chen\orcidA{}}
\affiliation{Key Laboratory of Nuclear Physics and Ion-beam Application (MoE), Institute of Modern Physics, Fudan University, Shanghai 200433, China}

\author{Zuo-Tang Liang\orcidB{}}
\affiliation{Key Laboratory of Particle Physics and Particle Irradiation (MOE),
Institute of Frontier and Interdisciplinary Science, Shandong University,
Qingdao, Shandong 266237, China}

\author{Yu-Gang Ma\orcidC{}}
\affiliation{Key Laboratory of Nuclear Physics and Ion-beam Application (MoE), Institute of Modern Physics, Fudan University, Shanghai 200433, China}

\author{Qun Wang\orcidD{}}
\affiliation{Peng Huanwu Center for Fundamental Theory and Department of Modern
Physics, University of Science and Technology of China, Hefei, Anhui
230026, China}

\maketitle

\paragraph{Introduction.}
Particles and fields are two fundamental forms of matter in our natural world. Particles are excitations or quanta of their corresponding fields, which are characterized by quantum numbers such as mass, charge, spin, and parity. Photons, for example, are quanta of electromagnetic fields that mediate electromagnetic forces among charged particles. In collisions at high energy scales, gluons appear as quanta of color fields that mediate interaction among quarks as elementary particles of strong interaction. At low energy scales, strong interaction is often characterized by mesons as effective degrees of freedom of quark and gluons, whose existence was proposed by Yukawa in 1935 in analogy with electromagnetic fields \citep{Yukawa:1935xg}. We now know nuclear forces as strong interaction at low energies have many components of meson fields~\citep{Sawada:1962,Manohar:1983md} $\sigma$, $\pi$, $\rho$, $\omega$, etc.. 
As the energy scale increases in nuclear reactions other meson fields carrying strangeness quantum number may come into play, such as $K$, $\phi$, etc.. As quanta of strong force fields, all these mesons with their specified quantum numbers were discovered before 1960s. 
Due to their short Compton wavelengths, experimentally it is much easier to detect particles than wave-like fields that are elusive and strongly fluctuate.

Recently, for the first time, the print of one kind of strong force field could be detected through the global spin alignment of vector mesons in heavy-ion collisions in the STAR experiment \citep{STAR:2022fan}. 
This is another breakthrough after the one in the measurement of the global spin polarization of $\Lambda$ in heavy-ion collisions \citep{STAR:2017ckg}. 
The global spin polarization arises from the initial orbital angular momentum (OAM) in nuclear collisions that is partially converted to the local OAM or vorticity leading to the polarization of hadrons through spin-orbit coupling in the interaction \citep{Liang:2004ph,Liang:2004xn,Wang}. The spin alignment of vector mesons in heavy-ion collisions along the OAM direction was first proposed by Liang and Wang some years ago \citep{Liang:2004xn}. 
The observable is the 00 element $\rho_{00}$ of the spin density matrix for vector mesons which can be measured 
by the angular distribution of their strong decay daughters in their rest frame \citep{Abelev:2008ag}.
The STAR data for $\rho_{00}$ show a surprisingly large positive deviation from 1/3, 
which is orders of magnitude larger than predictions by conventional mechanisms \citep{STAR:2022fan}.
In Ref. \citep{Sheng:2019kmk}, Sheng, Oliva and Wang proposed that
the local correlation or fluctuation of the $\phi$ meson field can produce
a large positive deviation for $\rho_{00}$ from 1/3 and thus provided a possible explanation.
Such $\phi$ field may originate from non-perturbative strong interaction coupled to $s$ and $\overline{s}$ 
and is connected with vacuum properties of quantum chromodynamics (QCD)\citep{Shuryak:1981ff,Shifman:1978bx}.

\begin{figure}
\includegraphics[width=4.0cm]{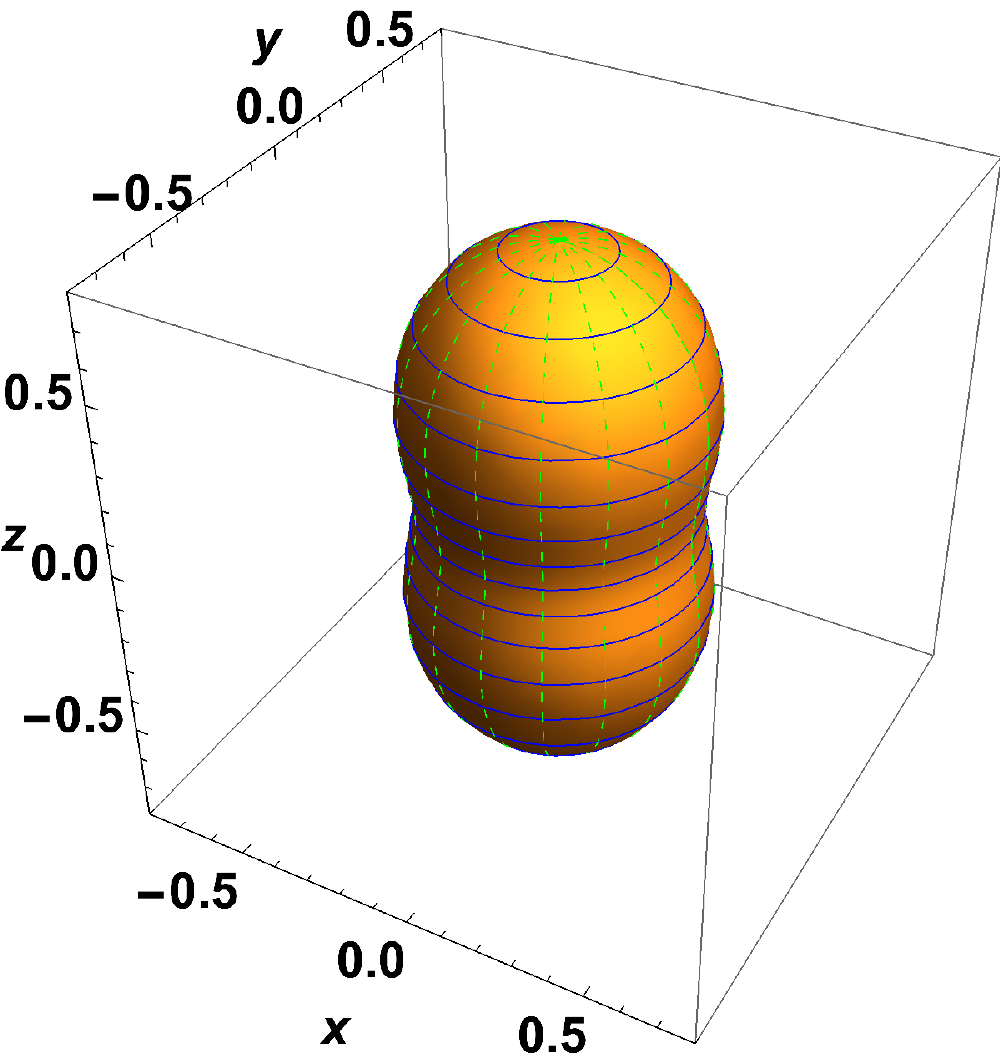}
\includegraphics[width=3.5cm]{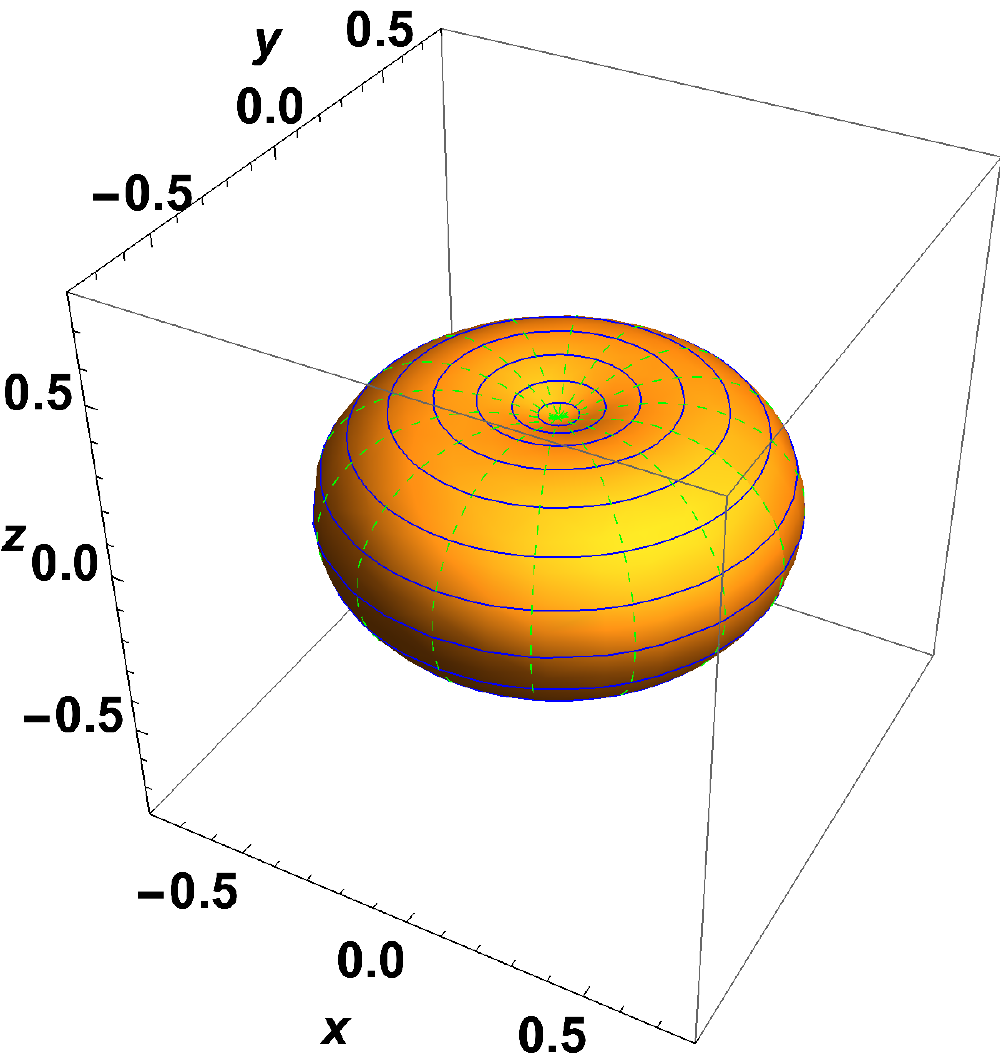}
\caption{\label{fig:rho-00}(Color online) The shape of $dN/d\mathrm{cos}\theta$ for the daughter particle in the vector meson's decay
with $\rho _{00}=1/2>1/3$ (left panel) and $\rho _{00}=1/6<1/3$ (right panel).}
\end{figure}

\paragraph{Vector meson spin alignment} 
The spin state of a system of spin-$S$ particles can be described by the spin density matrix $\hat\rho$, which is a $(2S+1)\times (2S+1)$ complex and Hermitian matrix for spin-$S$ particles with unit trace. 
The number of independent real variables in $\hat\rho$ is $4S(S+1)$. 
For example, for spin-1/2 particles, there are 3 independent real variables corresponding to a polarization vector $\mathbf{P}$ as in $\rho =(1/2)(1+\boldsymbol{\sigma}\cdot\mathbf{P})$
where $\boldsymbol{\sigma}$ are Pauli matrices. 
For spin-1 particles such as vector mesons, there are 8 independent real variables corresponding to a polarization vector $\mathbf{P}$ (3 variables) 
and a symmetric traceless tensor $T_{ij}$ (also called the tensor polarization, 5 variables). 
As a spin-1/2 particle, the $\Lambda$ hyperon's spin polarization can be measured through its parity-violating weak decay $\Lambda\rightarrow \mathrm{p}\pi^-$, 
since the preferred direction of the daughter proton's momentum is along its spin in the rest frame. 
However, this is not the case for vector mesons since they mainly decay through strong interaction in which the parity is conserved. 
So the elements of the spin density matrix that can be measured are its tensor components $T_{ij}$ and $\rho_{00}$ is associated with them. 
For the $\phi$ meson's strong decay $\phi\rightarrow K^+K^-$, the daughter particle's polar angular distribution is given by
\begin{equation}
\frac{dN}{d\mathrm{cos}\theta } = \frac{3}{4}\left[ 1-\rho_{00} + (3\rho_{00} -1 )\mathrm{cos}^2 \theta \right].
\label{eq:costheta}
\end{equation}
We see that $\rho _{00}$ appears in the coefficient of $\mathrm{cos}^2 \theta$ which can be determined by measuring the polar angle distribution. 
If $\rho _{00} =1/3$, the distribution is a constant indicating that the daughter particles are emitted isotropically. 
If $\rho _{00} \neq 1/3$, the probabilities of the vector meson in three spin states are not equal and then the emission of daughter particles is anisotropic. For $\rho _{00} \gtrless 1/3$, the polar angle distribution is in a cigar/discus shape (as shown in Fig. \ref{fig:rho-00}).  
Correspondingly, if $\rho _{00}>1/3$, the vector meson has a larger probability to be in the spin state $\lambda =0$ 
so that its average polarization vector (not the spin) is aligned to the spin quantization axis.  
If $\rho _{00}<1/3$, the vector meson has a larger probability to be in the spin states $\lambda =\pm 1$ and its average polarization vector is aligned in the plane perpendicular to the spin quantization axis. 
The quantity $\rho _{00}$ is thus referred to the spin alignment of the vector meson.

\paragraph{Experimental results.}
In Ref.~\cite{STAR:2022fan}, $\phi$ and $K^{*0}$ mesons are observed by paring of their decay daughters $K^{\pm}$ and $K\pi$, respectively, with subtraction of the combinatorial background and application of the so-called event mixing and rotation methods. Detail studies show that both techniques can effectively break the correlation between pairs in real events, and the yields of $\phi$ and $K^{*0}$ from two techniques are consistent~\cite{STAR:2022fan}. The spin quantization direction is chosen to be the normal direction of the second-order event plane constructed from charge particles collected by the STAR Time Projection Chamber (TPC). Then the polar angle distribution of Eq.~(\ref{eq:costheta}) is analyzed, and $\rho_{00}$ is extracted after correction for detection efficiency and acceptance. The spin quantization direction can be constructed by different detectors such as the shower maximum detector and the beam-beam counter, and they all give consistent results on the global spin alignment of $\phi$ and $K^{*0}$~\cite{STAR:2022fan}. In the following, we only discuss results with respect to the TPC event plane.

\begin{figure}
\includegraphics[width=7.0cm]{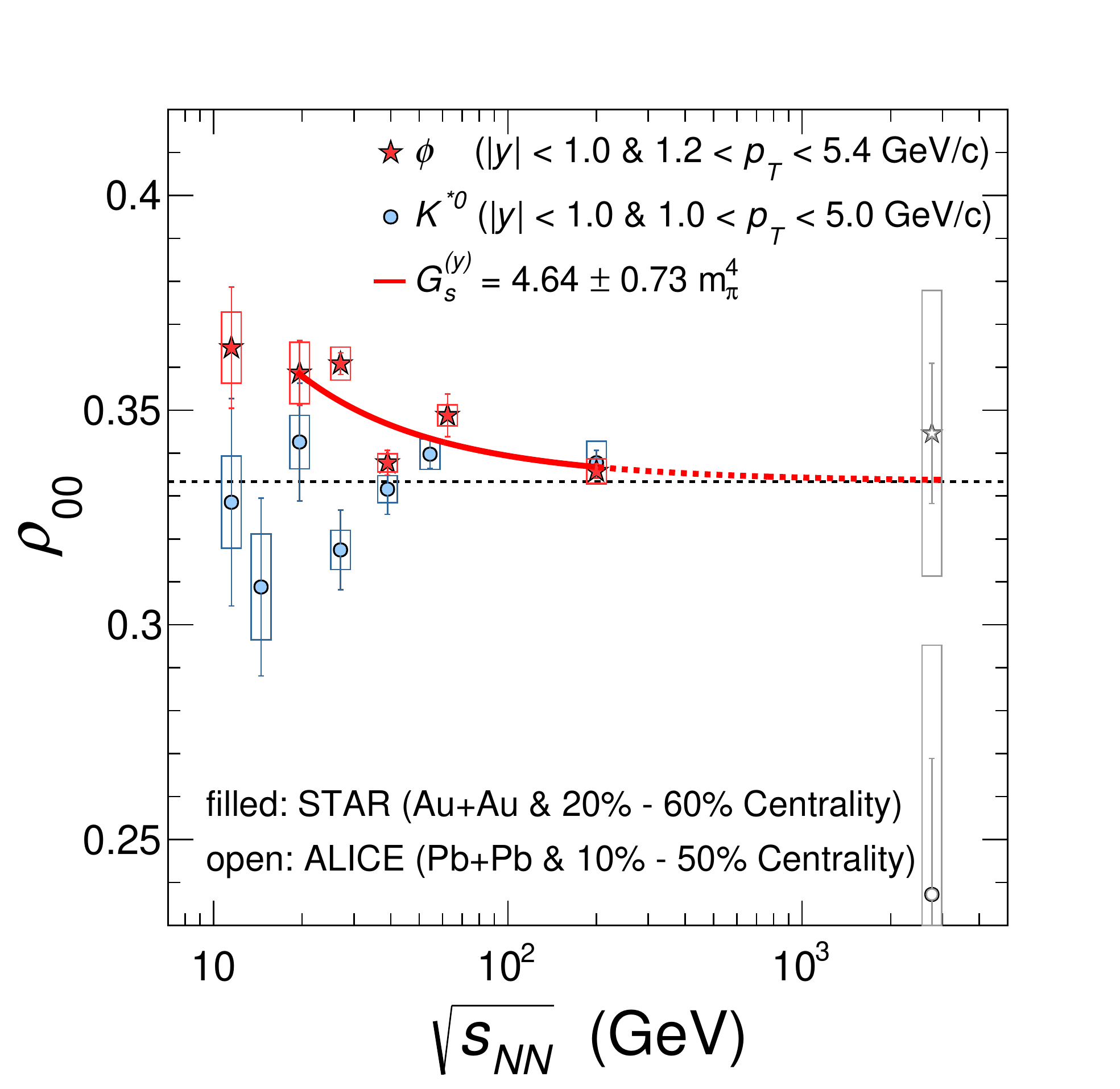}
\caption{\label{fig:rho00_data} (Color online) Measurements of $\rho_{00}$ 
with respect to the second-order event plane in high energy heavy-ion collisions. 
Stars represent the data for $\phi$ mesons~\cite{STAR:2022fan,Acharya:2019vpe}, circles represent the data for $K^{*0}$ mesons~\cite{STAR:2022fan,Acharya:2019vpe}.
The solid red line is a fit to data by the model based on strong force fields~\cite{Sheng:2019kmk}.}
\end{figure}

Figure~\ref{fig:rho00_data} shows the $\rho_{00}$ data for $\phi$ and $K^{*0}$ mesons in Au+Au collisions at $\mathrm{\sqrt{s_{NN}}}$= 11.5 to 200 GeV measured by the STAR collaboration~\cite{STAR:2022fan} from a dedicated Beam Energy Scan program and multiple years of high statistics event collection of Au+Au collisions at 200 GeV. The centrality categorizes events according to the observed number of tracks in each collision. Here the 0\% centrality corresponds to exactly head-on collisions, which produce the most tracks, while the 100\% centrality corresponds to barely glancing collisions, which produce the fewest tracks.
The STAR measurements presented in Fig.~\ref{fig:rho00_data} are for the centrality interval of  20\% to 60\%, where one expects the largest OAM among collisions and better signal to noise ratio in the experimental analysis. A complete set of results for centrality and transverse momentum dependence can be found in Ref. \cite{STAR:2022fan}.
The STAR's data give
\begin{eqnarray}
\rho_{00}^{\phi} &=& 0.3512 \pm 0.0017 (\mathrm{stat.}) \pm 0.0017 (\mathrm{syst.}), \nonumber\\
\rho_{00}^{K^{*0}} &=& 0.3356 \pm 0.0034 (\mathrm{stat.}) \pm 0.0043 (\mathrm{syst.}), \nonumber
\end{eqnarray}
which are obtained by averaging over results at energies from 11.5 to 62.4 GeV for $\phi$,
and from 11.5 to 54.4 GeV for $K^{*0}$~\cite{STAR:2022fan}.
Taking the total uncertainties as the sum in quadrature of statistical and systematic uncertainties, the results suggest that $\rho_{00}^\phi$ is above 1/3 with a significance of 7.4 $\sigma$, indicating a significant global spin alignment for the $\phi$ meson. The values of $\rho_{00}^{K^{*0}}$, however, are consistent with 1/3. Measurements in Pb+Pb collisions at $\mathrm{\sqrt{s_{NN}}}$= 2.76 TeV~\cite{Acharya:2019vpe} in the $p_T$ bin nearest to the mean $p_T$ for STAR data in the range from 1.0 to 5.0 GeV/c are also shown for comparison. They are consistent with the STAR data where magnitude of the spin alignment for both $\phi$ and $K^{*0}$ lie at 1/3 at the highest beam energies within large uncertainties.


\paragraph{Quark coalescence model.}
The quark coalescence model (QCM) is the convenient theoretical tool to describe the hadron production or hadronization in heavy-ion collisions. Therefore one can calculate the spin density matrix for vector mesons in terms of those of the quarks and antiquarks in QCM~\citep{Liang:2004xn,Sheng:2019kmk}. 
A non-relativistic QCM is the most simple and transparent one to this end~\citep{Sheng:2019kmk,Sheng:2023cpb}. 
In QCM, the meson's spin density operator can be constructed as $\hat\rho_M = \hat\rho_{q}\otimes\hat\rho_{\overline{q}}$, where $\hat\rho_{q}$ and $\hat\rho_{\overline{q}}$ are spin density operators of the quark and antiquark in spin and momentum space respectively. The elements $\rho_{\lambda_1\lambda_2}^{\mathrm{M}}(\mathbf{x},\mathbf{p})$ can be obtained by projecting $\rho_M$ onto two mesons' spin and momentum states and then taking a Fourier transformation with respect to the relative momentum of two states. 
The $\rho_{00}$ for the $\phi$ meson in phase space reads~\citep{Sheng:2019kmk}
\begin{eqnarray}
\rho_{00}^{\phi}(\mathbf{x},\mathbf{p})
&\approx & \frac{1}{3}-\frac{2}{3}\left\langle P_{q}^{y}\left(\mathbf{x}_1,\mathbf{p}_1\right)P_{\bar{q}}^{y}
\left(\mathbf{x}_2,\mathbf{p}_2\right)\right\rangle \nonumber\\
&&+\frac{2}{9}\left\langle \mathbf{P}_{q}\left(\mathbf{x}_1,\mathbf{p}_1\right)\cdot\mathbf{P}_{\bar{q}}
\left(\mathbf{x}_2,\mathbf{p}_2 \right)\right\rangle ,
\label{rho-00}
\end{eqnarray}
where $\mathbf{x}_1=\mathbf{x}+\Delta\mathbf{x}/2$, $\mathbf{x}_2=\mathbf{x}-\Delta\mathbf{x}/2$, $\mathbf{p}_1=\mathbf{p}/2+\Delta\mathbf{p}$, $\mathbf{p}_2=\mathbf{p}/2-\Delta\mathbf{p}$, 
the average is taken over $\Delta\mathbf{x}$ and $\Delta\mathbf{p}$ weighted by the $\phi$ meson's wave function, 
$\mathbf{P}_q(\mathbf{x}_1,\mathbf{p}_1)$ and $\mathbf{P}_{\overline{q}}(\mathbf{x}_2,\mathbf{p}_2)$ are spin polarization vectors of the quark and antiquark respectively,  
and the spin quantization direction is taken along the ${y}$ direction. 

Equation (\ref{rho-00}) clearly shows that the vector meson's spin alignment is determined by the local correlation of quark's and antiquark's polarization functions $\mathbf{P}_q(\mathbf{x}_1,\mathbf{p}_1)$ and $\mathbf{P}_{\overline{q}}(\mathbf{x}_2,\mathbf{p}_2)$ inside the phase space limited by the meson's wave function. 

\paragraph{Local correlation of $\phi$ fields.}
It is well known that fermions such as quarks at rest have magnetic moments proportional to their spins that are polarized along the direction of the magnetic field. For moving fermions, the electric field comes into play in the form of $\mathbf{p}\times\mathbf{E}$ (the spin-orbit coupling or spin-Hall effect). Similarly vorticity fields and vector meson fields can also polarize quarks and antiquarks. For $s$ and $\overline{s}$ that make the $\phi$ meson, the spin polarization vector is
\begin{eqnarray}
\mathbf{P}_{q/\overline{q}}&=&\frac{1}{2}\boldsymbol{\omega}+\frac{1}{2m_s}\boldsymbol{\varepsilon}\times{\bf p}\nonumber\\
&& \pm \frac{g_\phi}{2m_sT}{\bf B}_\phi \pm \frac{g_\phi}{2m_qE_pT}{\bf E}_\phi\times{\bf p},
\label{polar-all-dir}
\end{eqnarray}
where $T$ is the effective temperature of the quark matter when $s$ and $\overline{s}$ combine into the $\phi$ meson, $\boldsymbol{\varepsilon}$ and $\boldsymbol{\omega}$ are the electric and magnetic part of the thermal vorticity tensor, and $\mathbf{E}_\phi$ and $\mathbf{B}_\phi$ are the electric and magnetic part of the $\phi$ field, respectively.  Here we have neglected the effects from electromagnetic fields since they die away quickly in the late stage of the matter evolution in heavy-ion collisions. The effects from vorticity fields can also be neglected because their magnitudes from the measured $\Lambda$ polarization are too small to account for the observed $\rho_{00}^\phi$, but their terms are kept in Eq. (\ref{polar-all-dir}) just as a contrast to those of $\phi$ fields. The main difference between terms of vorticity fields and $\phi$ fields in Eq. (\ref{polar-all-dir}) is the sign for antiquarks: vorticity fields do not distinguish quarks from antiquarks while $\phi$ fields do (same as electromagnetic fields). 
We see in the following that this is essential to obtained a large vector meson spin alignment $\rho_{00}$ in the coalescence of $s$ and $\bar s$ into the $\phi$ meson.  

Substituting Eq. (\ref{polar-all-dir}) into Eq. (\ref{rho-00}), one obtains that $\rho_{00}^{\phi}$ depends on the local correlation of $\phi$ fields inside the $\phi$ meson's wave function.
A more rigorous approach based on the relativistic QCM has been formulated using the closed-time-path (Schwinger-Keldysh) method in quantum field theory \citep{Sheng:2022wsy, Sheng:2022ffb}. 
The result obtained for $\rho_{00}^{\phi}$ reads \citep{Sheng:2022wsy,Sheng:2022ffb}
\begin{eqnarray}
\rho_{00}^\phi (t,\mathbf{x},\mathbf{p})&\approx& \frac{1}{3}+C_1\left[\frac{1}{3}\boldsymbol\omega^\prime\cdot\boldsymbol\omega^\prime
-(\boldsymbol\epsilon_0\cdot\boldsymbol\omega^\prime)^2\right]\nonumber\\
&&+C_2\left[\frac{1}{3}\boldsymbol\varepsilon^\prime\cdot\boldsymbol\varepsilon^\prime
-(\boldsymbol\epsilon_0\cdot\boldsymbol\varepsilon^\prime)^2\right]\nonumber\\
&&-\frac{4g_\phi^2}{m_\phi^2T^2}\left\{ C_1\left[\frac{1}{3}{\bf B}_\phi^\prime\cdot{\bf B}_\phi^\prime-(\boldsymbol\epsilon_0\cdot{\bf B}_\phi^\prime)^2\right]\right.\nonumber\\
&&+\left.C_2\left[\frac{1}{3}{\bf E}_\phi^\prime\cdot{\bf E}_\phi^\prime-(\boldsymbol\epsilon_0\cdot{\bf E}_\phi^\prime)^2\right]\right\} ,
\label{result_coalescence}
\end{eqnarray}
where the fields with primes are in the $\phi$ meson's rest frame, and $C_1$ and $C_2$ are functions of $m_s$ (strange quark mass) and $m_\phi$ ($\phi$ meson mass).
We see that all terms appear in fields squared.
The momentum dependence of $\rho_{00}^\phi$ can be obtained by rewriting $\rho_{00}^\phi$ in terms of fields in the lab frame using Lorentz transformation.
By taking averages over space-time on the hadronization hyper-surface of the $\phi$ meson, one obtains $\rho_{00}^\phi$ as functions of momentum that can be compared with STAR's data \citep{STAR:2022fan}. 
The parameters are in the form of $\left\langle g_\phi^2 \mathbf{E}_{\phi}^2/T^2 \right\rangle$ and $\left\langle g_\phi^2 \mathbf{B}_{\phi}^2/T^2 \right\rangle$ and reflect local fluctuations of $\phi$ fields~\citep{Sheng:2022wsy}. 
If we assume that local field fluctuations are different in the transverse (labeled as $i$) and longitudinal direction (labeled as $z$) with respect to the beam direction $z$ in heavy-ion collisions, then $\rho_{00}^\phi$ depends on two parameters $F_T^2\equiv \langle g_\phi^2 B_{\phi,i}^2/T^2 \rangle = \langle g_\phi^2 E_{\phi,i}^2/T^2 \rangle$ and $F_z^2\equiv \langle g_\phi^2 B_{\phi,z}^2/T^2 \rangle = \langle g_\phi^2 E_{\phi,z}^2/T^2\rangle$. The values of two parameters are determined by fitting STAR's data on $\rho_{00}$ as functions of collision energies in Fig. \ref{fig:rho00_data}. With fitted values of two parameters one can predict the transverse momentum spectra of $\rho_{00}^\phi$ at all available collision energies which are in good agreement with STAR's data~\citep{STAR:2022fan}.

\paragraph{Summary and outlook}

An unexpected large global spin alignment of $\phi$ mesons has been observed by the STAR Collaboration~\citep{STAR:2022fan} in relativistic heavy-ion collisions.
By using the quark coalescence model for hadron production, Refs.~\cite{Sheng:2019kmk,Sheng:2022wsy,Sheng:2022ffb} provide a good interpretation of the data~\citep{STAR:2022fan}.   
According to this interpretation, such a large global spin alignment of $\phi$ mesons is induced by the local correlation or fluctuation of strong force fields on the hadronization hyper-surface.
The average values $\langle \mathbf{E}_{\phi}^2 \rangle$ and $\langle \mathbf{B}_{\phi}^2\rangle$ reflect the local fluctuation of $\phi$ fields and are expected to be calculable using lattice QCD. 
These studies open a new avenue to investigate properties of the strongly interacting quark matter as well as non-perturbative properties of strong interaction. 
In addition, relativistic heavy-ion collisions are usually called ``small bang'' in contrast to the Big Bang of the universe. In such an analog, hadronization on the freeze-out hyper-surface corresponds to the stage of the early universe in which particles are decoupled from the interaction during the Big Bang. The vector meson's spin alignment is similar to polarization modes of cosmic microwave background radiation.  

We note that such an explanation is still subject to debate and further verification. More systematical studies in both experiments and theories need to be done to clarify the deep physics behind the phenomena.

\paragraph{Conflict of interest. }

The authors declare that they have no conflict of interest.

\paragraph{Acknowledgements.}
The authors thank A.H. Tang for insightful discussion. 
This work was supported in part by the National Key Research and Development Program of China under Contract No. 2022YFA1604900, 
by the National Natural Science Foundation of China (NSFC) under Contract Nos. 11890710, 11890713, 11890714, 12025501, 12135011 and 12147101,  
and by the Strategic Priority Research Program of the Chinese Academy of Sciences (CAS) under Grant No. XDB34030102.

\bibliographystyle{unsrt}
\bibliography{ref-vm2}

\end{document}